\documentclass[a4paper,12pt]{article}
\usepackage{amssymb}
\usepackage{amsmath}
\usepackage{epsfig}

\usepackage{graphics}

\usepackage{latexsym}
\usepackage{rotating}
\title{The Kac-Wakimoto Equation Is Not Integrable}

\author {
Asl{\i} Pekcan\thanks{Email:asli@fen.bilkent.edu.tr} \\
{\small Department of Mathematics, Faculty of Science} \\
{\small Hacettepe University, 06800 Ankara - Turkey}\\
}

\date{\nonumber}
\begin{document}
\maketitle
\date{\nonumber}

\baselineskip 17pt

\numberwithin{equation}{section}

\begin{abstract} We study the $(3+1)$-dimensional eight-order nonlinear wave equation associated with the
principal representation of the exceptional affine Lie algebra $E_6^{(1)}$, which was constructed by Kac and Wakimoto and stated that $N$-soliton solution of the equation can be formulated. We show that the equation is not Hirota integrable since it does not have three-soliton solution, even it has one- and two-soliton solutions.
\end{abstract}

\newtheorem{thm}{Theorem}[section]
\newtheorem{Le}{Lemma}[section]
\newtheorem{defi}{Definition}[section]
\newtheorem{ex}{Example}[section]
\newtheorem{pro}{Proposition}[section]

\section{Introduction}

Integrable higher-dimensional nonlinear partial differential equations are used to model problems
in physics, particularly in nonlinear optics, hydrodynamics, and plasma physics. Even though there are many $(1+1)$-dimensional integrable equations, there are a few higher-dimensional integrable equations known \cite{Hietarinta2}-\cite{Hietarinta6}. Date, Jimbo, Kashiwara, and Miwa studied the connection between the soliton theory and the classical affine Kac-Moody algebras by using the boson-fermion correspondence in $2$-dimensional quantum field theory \cite{DJKM1}-\cite{DJKM4}. By considering the groups $GL_{\infty}$, $SL_n$, $O_{\infty}$, they construct the KP, KdV, Boussinesq, and BKP hierarchies and also $N$-soliton solutions of these hierarchies by iterated application of vertex operators to the trivial solution $1$. In \cite{KW}, Kac and Wakimoto search for an answer about the general way of constructing the hierarchies associated to arbitrary loop groups, including the exceptional ones. In their paper, they construct a hierarchy, particularly, a $(3+1)$-dimensional eight-order nonlinear partial differential equation associated with
the exceptional affine Lie algebra $E_6^{(1)}$. Kac and Wakimoto claims that the $N$-soliton solutions of all the hierarchies constructed in \cite{KW}, can be written by iterated application of vertex operators to $1$ as in the works \cite{DJKM1}-\cite{DJKM4}. This $(3+1)$-dimensional eight-order nonlinear equation has the following Hirota bilinear form \cite{KW}
\begin{equation}\label{HirotabilinearKW}
P(D)\{f\cdot f\}=(D_x^8+a D_x^3D_y+b D_z^2+c D_xD_t)\{f\cdot f\}=0,
\end{equation}
where
\begin{equation}\label{abc}
a=-280\sqrt{6},\, b=210,\, c=-240\sqrt{2}.
\end{equation}
Here $D$ is the Hirota $D$-operator given by
\begin{align}
[D_{x}^{m_1}D_t^{m_2}\ldots]\{f\cdot g\}=[(\partial_x-\partial_{x'})^{m_1}
(\partial_t-\partial_{t'})^{m_2}\ldots]f(x,t,\ldots)\cdot g(x',t',\ldots)|_{x'=x,t'=t,\ldots}
\end{align}
where $m_i$, $i=1,2,\ldots$ are positive integers and $x,t,\ldots$ independent variables. In this letter, we will call
the equation (\ref{HirotabilinearKW}) as the Kac-Wakimoto (KW) equation.
In \cite{DODD}, Dodd finds one- and two-soliton solutions of the KW equation by following the Hirota direct method \cite{Hirota}, \cite{Hietarinta1} and
states that $N$-soliton solution can also be calculated similarly, which is indeed wrong.

To the best of our knowledge, there has been no Lax pair found for the KW equation.
For investigating the integrability of the KW equation, Sakovich considers the Painlev\'{e} analysis \cite{SAKOVICH}.
He proves that the equation (\ref{HirotabilinearKW}) obtained by the transformation $u=2\partial_x f(x, t, y, z)$,
\begin{align}
&u_{8x}+28u_xu_{6x}+28u_{xx}u_{5x}+70u_{xxx}u_{4x}+210u_x^2u_{4x}+420u_xu_{xx}u_{xxx}+420u_x^3u_{xx}\nonumber\\
&+a(u_{xxxy}+3u_yu_{xx}+3u_xu_{xy})+bu_{zz}+cu_{xt}=0,
\end{align}
where $a, b, c$ are as in (\ref{abc}), fails the Painlev\'{e} test. Indeed, this result does not depend on the values of
$a$, $b$, and $c$; i.e. the conclusion on non-integrability of the KW equation is also valid for the reductions of this equation, including
the KW equation associated with the exceptional affine Lie algebra $D_4^{(3)}$ which is the $z$-independent reduction of (\ref{HirotabilinearKW}).

\section{Soliton Solutions}

The KW equation (\ref{HirotabilinearKW}) can be written in standard form by using the transformations $u=2\partial_x \ln f(x, t, y, z)$ or
$u=2\partial_x^2 \ln f(x, t, y, z)$. Let us take the latter one. The equation (\ref{HirotabilinearKW}) becomes
\begin{equation}
u_{8x}+35u_{4x}^2+28u_{xx}u_{6x}+210u_{xx}^2u_{4x}+105u_{2x}^4-280\sqrt{6}(u_{xxxy}+3u_{xx}u_{xy})+210u_{zz}-240\sqrt{2}u_{xt}=0.
\end{equation}
Now we find one- and two-soliton solutions of the KW equation \cite{DODD}. For one-soliton solution, we take the ansatz $f=1+\varepsilon e^{\theta_1}$, where $\theta_1=a_1x+b_1y+c_1z+d_1t+\delta_1$. Here $a_1, b_1, c_1, d_1, \delta_1$ are constants. By inserting the ansatz into (\ref{HirotabilinearKW})
and making the powers of $\varepsilon$ equal to zero, we obtain the dispersion relation $P(\vec{p}_1)=0$, $\vec{p}_1=(a_1,b_1,c_1,d_1)$, that is
\begin{equation}\label{dispersion}
a_1^8-280\sqrt{6}a_1^3b_1+210c_1^2-240\sqrt{2}a_1d_1=0.
\end{equation}
By the transformation $u=2\partial_x^2\ln f(x, t, y, z)$ and setting $\varepsilon=1$, we get one-soliton solution of the KW equation,
\begin{equation}
\displaystyle u(x,t,y,z)=(a_1^2/2)\mathrm{sech}^2(\theta_1),
\end{equation}
where $\theta_1=a_1x+b_1y+c_1z+d_1t+\delta_1$ with the dispersion relation (\ref{dispersion}) satisfied.

For two-soliton solution, take the ansatz $f=1+\varepsilon(e^{\theta_1}+e^{\theta_2})+\varepsilon^2 S_{12}e^{\theta_{12}}$, where $\theta_i=a_ix+b_iy+c_iz+d_it+\delta_i$, $i=1, 2$, and
$\theta_{12}=a_{12}x+b_{12}y+c_{12}z+d_{12}z+\delta_{12}$. If we insert the ansatz into (\ref{HirotabilinearKW}), the coefficient of $\varepsilon$ gives that the parameters of $\theta_i$, $i=1, 2$, satisfy the dispersion relation
\begin{equation}\label{dispersion2ss}
P(\vec{p}_i)=a_i^8-280\sqrt{6}a_i^3b_i+210c_i^2-240\sqrt{2}a_id_i=0, i=1, 2.
\end{equation}
The coefficient of $\varepsilon^3$ gives
\begin{equation}
P(\vec{p}_1-\vec{p}_{12})e^{\theta_1+\theta_{12}}+P(\vec{p}_2-\vec{p}_{12})e^{\theta_2+\theta_{12}}=0,
\end{equation}
where $\vec{p}_{12}=(a_{12},b_{12},c_{12},d_{12})$. Therefore we obtain the relations $P(\vec{p}_i-\vec{p}_{12})=0$, $i=1, 2$; i.e.,
\begin{equation}\label{epsilon^3}
(a_i-a_{12})^8-280\sqrt{6}(a_i-a_{12})(b_i-b_{12})+210(c_i-c_{12})^2-240\sqrt{2}(a_i-a_{12})(d_i-d_{12})=0, i=1, 2.
\end{equation}
\noindent From the coefficient of $\varepsilon^2$, we have
\begin{align}\label{epsilon^2}
&S_{12}P(D)\{1\cdot e^{\theta_{12}}+e^{\theta_{12}}\cdot 1\}+P(D)\{(e^{\theta_1}+e^{\theta_2})\cdot (e^{\theta_1}+e^{\theta_2})\}\nonumber\\
&=2S_{12}P(\partial)e^{\theta_{12}}+2P(\vec{p}_1-\vec{p}_2)e^{\theta_1+\theta_2}=0.
\end{align}

\noindent In \cite{DODD}, it is stated that if (\ref{epsilon^3}) holds then two-soliton solution of (\ref{HirotabilinearKW}) is given
by $f=1+\varepsilon(e^{\theta_1}+e^{\theta_2})+\varepsilon^2 S_{12}e^{\theta_{12}}$ with
\begin{equation*}
S_{12}=P_{12}/Q_{12},
\end{equation*}
where with $a_{-}=a_1-a_2$, $b_{-}=b_1-b_2$, $c_{-}=c_1-c_2$, and $d_{-}=d_1-d_2$,
\begin{align*}
&P_{12}=-(a_{-}^8-280\sqrt{6}a_{-}^3b_{-}+210c_{-}^2-240\sqrt{2}a_{-} d_{-})\\
&Q_{12}=a_{12}^8-280\sqrt{6}a_{12}^3b_{12}+210c_{12}^2-240\sqrt{2}a_{12}d_{12}.
\end{align*}
It is also noted that the equation (\ref{epsilon^3}) is satisfied if we take $(a_{12}=a_1+a_2, b_{12}=b_1+b_2, c_{12}=c_1+c_2, d_{12}=d_1+d_2)$, that is
$\theta_{12}=\theta_1+\theta_2$. But according to Dodd, there are many other possible solutions. A set of solutions to (\ref{epsilon^3}) is also given:
\begin{equation}
b_{12}=(p_1m_2-p_2m_1)/(p_2n_1-p_1n_2),\quad d_{12}=(m_1n_2-n_1m_2)/(p_2n_1-p_1n_2),
\end{equation}
where with $\alpha_i=a_i-a_{12}$, and $\gamma_i=c_i-c_{12}$,
\begin{equation}
m_i=\alpha_i^8-280\sqrt{6}\alpha_i^3b_i+210\gamma_i^2-240\sqrt{2}\alpha_id_i,\, n_i=280\sqrt{6}\alpha_i^3,\, p_i=240\sqrt{2}\alpha_i,\, i=1, 2.
\end{equation}
In addition to that one numerical example considering the case when $\theta_{12}\neq \theta_1+\theta_2$ is given. But this is clearly a mistake.
Obviously, the equation (\ref{epsilon^2}) is only satisfied if $\theta_{12}=\theta_1+\theta_2$ which gives
\begin{equation}\label{A12}
\displaystyle S_{12}=P_{12}/Q_{12}=-P(\vec{p}_1-\vec{p}_2)/P(\vec{p}_1+\vec{p}_2).
\end{equation}
Thus by setting $\varepsilon=1$, we get two-soliton solution of the KW equation
\begin{equation}
\displaystyle u=\frac{2\{(S_{12}(a_1+a_2)^2+(a_1-a_2)^2)e^{\theta_1+\theta_2}+a_1^2e^{\theta_1}(1+S_{12}e^{2\theta_2})+a_2^2e^{\theta_2}(1+S_{12}e^{2\theta_1})    \}}
{(1+e^{\theta_1}+e^{\theta_2}+S_{12}e^{\theta_{12}})^2},
\end{equation}
where $S_{12}$ is given by (\ref{A12}), and $\theta_i=a_ix+b_iy+c_iz+d_it+\delta_i$, $i=1, 2$ with the dispersion relation (\ref{dispersion2ss}) satisfied.

To obtain three-soliton solution of the KW equation, we insert the ansatz $f=1+\varepsilon(e^{\theta_1}+e^{\theta_2}+e^{\theta_3})+\varepsilon^2(S_{12}e^{\theta_1+\theta_2}
+S_{13}e^{\theta_1+\theta_3}+S_{23}e^{\theta_2+\theta_3})+\varepsilon^3 S_{123}e^{\theta_1+\theta_2+\theta_3}$, where $\theta_i=a_ix+b_iy+c_iz+d_it+\delta_i$, $i=1, 2, 3$, into (\ref{HirotabilinearKW}) and make the coefficients of the powers of $\varepsilon$ equal to zero. The coefficient of $\varepsilon$ gives that the parameters of $\theta_i$, $i=1, 2, 3$, satisfy the dispersion relation $P(\vec{p}_i)=0$ that is
\begin{equation}\label{dispersion3ss}
a_i^8-280\sqrt{6}a_i^3b_i+210c_i^2-240\sqrt{2}a_id_i=0, i=1, 2, 3.
\end{equation}
From the coefficient of $\varepsilon^2$, we obtain the constants $S_{ij}$,
\begin{equation}
\displaystyle S_{ij}=-P(\vec{p}_i-\vec{p}_j)/P(\vec{p}_i+\vec{p}_j), 1\leq i < j \leq 3.
\end{equation}
The coefficients of $\varepsilon^3$ and $\varepsilon^4$ give two expressions for $S_{123}$ to be satisfied. To be consistent, these expressions must be equal to each other. Thus we get a condition which is called as the three-soliton solution condition (3SC) \cite{Hietarinta1}
\begin{align}\label{3ssC}
(3SC)&=P(\vec{p}_1+\vec{p}_2+\vec{p}_3)P(\vec{p}_1-\vec{p}_2)P(\vec{p}_1-\vec{p}_3)P(\vec{p}_2-\vec{p}_3)\nonumber \\
&+P(\vec{p}_3-\vec{p}_1-\vec{p}_2)P(\vec{p}_1-\vec{p}_2)P(\vec{p}_1+\vec{p}_3)P(\vec{p}_2+\vec{p}_3)\nonumber \\
&+P(\vec{p}_1-\vec{p}_2-\vec{p}_3)P(\vec{p}_1+\vec{p}_2)P(\vec{p}_1+\vec{p}_3)P(\vec{p}_2-\vec{p}_3)\nonumber \\
&+P(\vec{p}_2-\vec{p}_1-\vec{p}_3)P(\vec{p}_1+\vec{p}_2)P(\vec{p}_1-\vec{p}_3)P(\vec{p}_2+\vec{p}_3)=0
\end{align}
with the dispersion relation $P(\vec{p}_i)=0$, $i=1, 2, 3$, must be satisfied.
\begin{thm}
The KW equation (\ref{HirotabilinearKW}) does not possess three-soliton solution.
\end{thm}
\textbf{Proof.} A nonlinear partial differential equation possesses three-soliton solution if (3SC) given by (\ref{3ssC}) is satisfied directly; without any additional condition on the parameters except the dispersion relation.
Therefore, to prove the non-existence of three-soliton solution for the KW equation, it is enough to give a particular example so that (3SC) is not
satisfied. Choose the parameters $a_i$, $c_i$, and $d_i$ as
\begin{eqnarray}\displaystyle
&& a_1=1,\quad c_1=1/3, \quad d_1=-1/2, \nonumber\\
&& a_2=-1,\quad c_2=2/3, \quad d_2=2,  \nonumber\\
&& a_3=1/2,\quad c_3=1, \quad d_3=3/4.
\end{eqnarray}
By using the dispersion relation (\ref{dispersion3ss}), we deduce the parameters $b_i$, $i=1, 2, 3$ as
\begin{equation}
\displaystyle b_1=(9c-2b-18)/18a,\quad  b_2=(-18c+4b+9)/9a,\quad  b_3=-(96c+256b+1)/32a,
\end{equation}
 where $a, b,$ and $c$ are as in (\ref{abc}). We obtain the followings:
\begin{align}
& P(\vec{p}_1-\vec{p}_2)=15c-13b/3+240,\quad P(\vec{p}_1+\vec{p}_2)=b, \nonumber\\
&P(\vec{p}_1-\vec{p}_3)=-3c/16+103b/72-15/128,\quad  P(\vec{p}_1+\vec{p}_3)=-129c/16-1843b/72+2835/128,\nonumber\\
&P(\vec{p}_2-\vec{p}_3)=-21c/4-511b/18+2835/128,\quad  P(\vec{p}_2+\vec{p}_3)=-3c/4+67b/18-15/128,\nonumber\\
&P(\vec{p}_3-\vec{p}_1-\vec{p}_2)=-9c/16-25b/24,\quad  P(\vec{p}_2-\vec{p}_1-\vec{p}_3)=-195c/16-9593b/72+95625/64\nonumber\\
&P(\vec{p}_1-\vec{p}_2-\vec{p}_3)=219c/16+1937b/72+1215/64,\quad  P(\vec{p}_1+\vec{p}_2+\vec{p}_3)=9c/16+73b/24.\nonumber
\end{align}
Hence the terms in (3SC) become
\begin{equation}
P(\vec{p}_1+\vec{p}_2+\vec{p}_3)P(\vec{p}_1-\vec{p}_2)P(\vec{p}_1-\vec{p}_3)P(\vec{p}_2-\vec{p}_3)=(4375/8192)(A_1\sqrt{2}+B_1/36),
\end{equation}
with $A_1=(103)(2099)(7297)(4451)$ and $B_1=-(653)(144470533189)$,
\begin{equation}
P(\vec{p}_3-\vec{p}_1-\vec{p}_2)P(\vec{p}_1-\vec{p}_2)P(\vec{p}_1+\vec{p}_3)P(\vec{p}_2+\vec{p}_3)=(625/8192)(A_2\sqrt{2}+B_2/36),
\end{equation}
with $A_2=-(37)(27817)(36723949)$ and $B_2=(5)(7)(70351)(698213011)$,
\begin{equation}
P(\vec{p}_1-\vec{p}_2-\vec{p}_3)P(\vec{p}_1+\vec{p}_2)P(\vec{p}_1+\vec{p}_3)P(\vec{p}_2-\vec{p}_3)=(30625/8192)(A_3\sqrt{2}+B_3/576),
\end{equation}
with $A_3=-(3)(9749)(3671)(117119)$ and $B_3=(7)(157)(15534667)(6210623)$, and
\begin{equation}
P(\vec{p}_2-\vec{p}_1-\vec{p}_3)P(\vec{p}_1+\vec{p}_2)P(\vec{p}_1-\vec{p}_3)P(\vec{p}_2+\vec{p}_3)=-(4375/8192)(A_4\sqrt{2}+B_4/576),
\end{equation}
with $A_4=(3)(5)(567881)(75209)$ and $B_4=(7)(367)(546569)(987697)$. When we insert the above values into (3SC), we get a nonzero value which is
$\displaystyle (3SC)=(1875/32)(A_5\sqrt{2}+B_5/256)$, where $A_5=-(3^3)(5^2)(11)(106846721)$ and $B_5=(7)(37)(563)(2090660753)$. Hence (3SC) is not satisfied for the KW equation, which yields that the KW equation
does not possess three-soliton solution. $\Box$

Note that the $t$-, $y$-, or $z$-independent reductions of the KW equation also do not satisfy (3SC) directly. Thus, the KW equation given by (\ref{HirotabilinearKW}) and its $t$-, $y$-, or $z$-independent reductions  are not integrable in Hirota sense.

\section{Conclusion}
 We have studied the $(3+1)$-dimensional eight-order KW equation constructed by Kac and Wakimoto associated with the exceptional affine Lie algebra $E_6^{(1)}$. Due to construction, Kac and Wakimoto claims that $N$-soliton solution of the KW equation can be formulated by iterated application of
 vertex operators to the trivial soliton $1$. We obtain one- and two-soliton solutions of the KW equation. We mention about the mistake in Dodd's paper related to two-soliton solution of the KW equation. We check whether the KW equation satisfy the three-soliton solution condition directly. We give specific values to the parameters of the three-soliton solution ansatz and conclude that the condition is not satisfied; i.e., the KW equation is not integrable in Hirota sense.

\section{Acknowledgment}
  This work is partially supported by the Scientific
and Technological Research Council of Turkey (T\"{U}B\.{I}TAK).\\

\end{document}